\documentclass{PoS}
\let\ifpdf\relax
\usepackage[]{graphicx}
\usepackage[authoryear]{natbib} 
\bibpunct{(}{)}{;}{a}{}{,}

\title{Stacking of SKA data: comparing uv-plane
and and image-plane stacking}

\ShortTitle{Stacking SKA:  uv- vs image-plane stacking}

\author{\speaker{Kirsten K.\ Knudsen}$^1$, 
Lukas Lindroos$^1$, Wouter Vlemmings$^1$, 
John Conway$^1$, Iv\'an Mart\'i-Vidal$^1$\\
	$^1$Chalmers University of Technology \\
        E-mail: \email{kirsten.knudsen@chalmers.se},
\email{lindroos@chalmers.se}}

\abstract{
Stacking as a tool for studying objects that are not individually detected 
is becoming popular even for radio interferometric data, and will be
widely used in the SKA era.  Stacking is
typically done using imaged data rather than directly using the
visibilities (the {\it uv}-data).  
We have investigated and developed a novel algorithm to do stacking
using the uv-data.  We have performed extensive simulations comparing to
image-stacking, and summarize the results of these simulations.  
Furthermore, we disuss the implications in light of the vast data volume produced by the SKA. 
Having access to the uv-stacked data provides a great
advantage, as it allows the possibility to properly analyse the result with
respect to calibration artifacts as well as source properties such as size.  
For SKA the main challenge lies in archiving the uv-data. 
For purposes of robust stacking analysis,
it would be strongly desirable to either keep the calibrated uv-data at least
in an average form, or implement a {\it stacking queue} where stacking
positions could be provided prior to the observations and the uv-stacking is
done almost in real time.  
}

\FullConference{
Advancing Astrophysics with the Square Kilometre Array\\
June 8-13, 2014\\
Giardini Naxos, Italy}

\newcommand{\skipthis}[1]{}

\begin{document}

\section{Introduction}

Stacking is known as a tool to average together data for a given set of
objects.  In most cases, this is done for imaging and photometry where the sources in
question are not individually detected (though have known positions), 
In that case, stacking the data yields an
average detection or upper limit.  This is done for data across the whole
electromagnetic spectrum ranging from X-rays to radio 
\citep[e.g.][]{nandra02,worsley05,knudsen05,dole06,carilli08}.  

A large part of the imaging produced in radio comes from interferometric
observations, with many large surveys available or planned (e.g., VLA-COSMOS,
\citealt{schinnerer07}, and the VLA 1.4 GHz Survey of the Extended Chandra Deep
Field South, \citealt{miller13}).
The success of stacking using the interferometric images depends on how well
the image reconstruction and deconvolution process has been.  For example, as
interferometric observations act as a spatial filter, flux could be missing
or very bright sources can have caused increased noise and/or artifacts in
the imaging.

The SKA will offer access to so much data and large surveys, that stacking
will become a frequently used tool to study the fainter sources.  This
naturally gives rise to the question, what are efficient stacking tools?  
We have investigated stacking directly in the uv-plane on the visibilities
rather than in the image-plane.  We have developed a novel algorithm and
compared this to image-stacking.  In this chapter we discuss some of the
implications of the results for stacking of SKA data.  

While the scientific motivation for our work is studying faint, high-redshift
galaxies with star formation rates below a few M$_\odot$/yr, galaxies, the
algorithm analysed here would also find application in other fields of
astronomy, and both for spectral line and continuum studies.  

\section{A novel {\it uv}-stacking algorithm}

We have developed a new stacking algorithm for radio inteferometric data.
The algorithm works directly on the uv-data and is applicable to any radio
interferometric data.  The algorithm has been tested on simulated data
mimicking JVLA and ALMA observations and compared to image-stacking, i.e.
stacking on the reconstructed and deconvolved images.  This is presented in
the paper \citet{lindroos15}.

In summary, the algorithm works as follows:  For sources (i.e. given positions) 
within a single pointing the visibilities are recalculated using 
\begin{equation}
   V_\mathrm{stack}(u,v,w) = V(u,v,w) \frac{\Sigma_{k=1}^N w_k
\frac{1}{A_N(\mathbf{\hat{S}_k})} e^{\frac{2\pi}{\lambda} i
\mathbf{B}\cdot\left( \mathbf{\hat{S}_0} - \mathbf{\hat{S}_k} \right)} }{\Sigma_{k=1}^N w_k}
   \label{eq:uvstack}
\end{equation}
where 
$\mathbf{\hat{S}_0}$ is the unit vector pointing to the phase centre, 
$\mathbf{\hat{S}_k}$ is the unit vector pointing to the stacking positions, 
$A_N(\mathbf{\hat{S}_k})$ describes the primary beam attenuation in the
direction $\mathbf{\hat{S}_k}$, 
$\mathbf{B}$ is the baseline of the visibility, $\lambda$ is the wavelength,
and $w_k$ is the weight of the stacking position.   This means, that the
visibilities are not duplicated, yielding the important advantage that the
size of the data set is not increased and kept managable. 
Furthermore, as the computation for each visilibity is done independently, the
code can be parallellized and thus run quickly for large sets. 

The whole algorithm is designed to be able to handle the effects from e.g. mosaics
and wide-field observations.  For mosaics, the algorithm is run for each pointing
individually, and subsequently concatenated into one data set.  Weights are
recalculated to include the relative weights between pointings.  

In the design of the algorithm, wide-field effects (for example, stretching of the {\it uv}-coordinates local to the stacked position) do not pose any
restrictions on point-sources.  In the case of extended sources, we estimate
that the baseline re-projection effects should not be larger than at most a
few per cent in the recovered size when using the JVLA, though larger for the longer SKA-mid baselines.

In order to test the algorithm, we have carried out extensive
simulations of JVLA and ALMA type of data.  These data represent both the GHz
and 100's GHz regime.  
The JVLA like simulations provide the closest representation to the SKA data,
in particular SKA-MID. 
The simulations have systematically covered
different aspects.  The sources input to the simulations were defined as
'bright foreground sources' and 'target sources'.  The former represent the
bright sources across the sky and typically contribute to the noise.  
The latter are the sources of interest to be stacked.  
We have investigated several effects, in particular:  sub-pixel sampling,
wide-field effects, bright foreground sources, extended sources, and  
mosaiced fields.  Below we focus on the extended foreground and target sources.

\subsection{Extended bright foreground sources}

The presence of bright sources in the data impacts on the dynamic range.
While this is already a challenge for present-day interferometric arrays such
as the JVLA, this will be even more pronounced for the SKA.   

Bright source present in data need to be removed, and this is
generally the case both for image- and uv-stacking.   Typically, bright
sources are removed using the best available model, however, there will always be
residuals left depending on the depth of the deconvolution, the quality of
the data, and how well the sources can be modeled.  In the GHz range, many
radio sources have complex morphology, e.g. jets and lopes of radio
galaxies, and this makes it difficult to completely model the flux distribution.  

We have run extensive simulations, where bright, extended foreground sources were
introduced.  The size of the sources varied between a FWHM of 0.5" to 5" and
the flux varied following a log-polynomial distribution derived from the
COSMOS field \citep{bondi08}.  After imaging (using Multi-Frequency Synthesis, 
\citealt{conway90}, combined with {\it w}-projection, \citealt{cornwell08}) and
deconvolution (using CLEAN, \citealt{hogbom74}), a residual measurement set is
produced and the noise in the centre of the deconvolved image is twice that
of the thermal noise limit.  Simulations were run in a Monte Carlo setup with
typically 100 realisations to reduce statistical variations.  Here we focus
on the setup of a JVLA A-array configuration with a central frequency of
1.4\,GHz and a bandwidth of 250\,MHz (for further
details, see \citealt{lindroos15}).  

\begin{figure}
\begin{center}
\includegraphics[width=8cm,height=8cm]{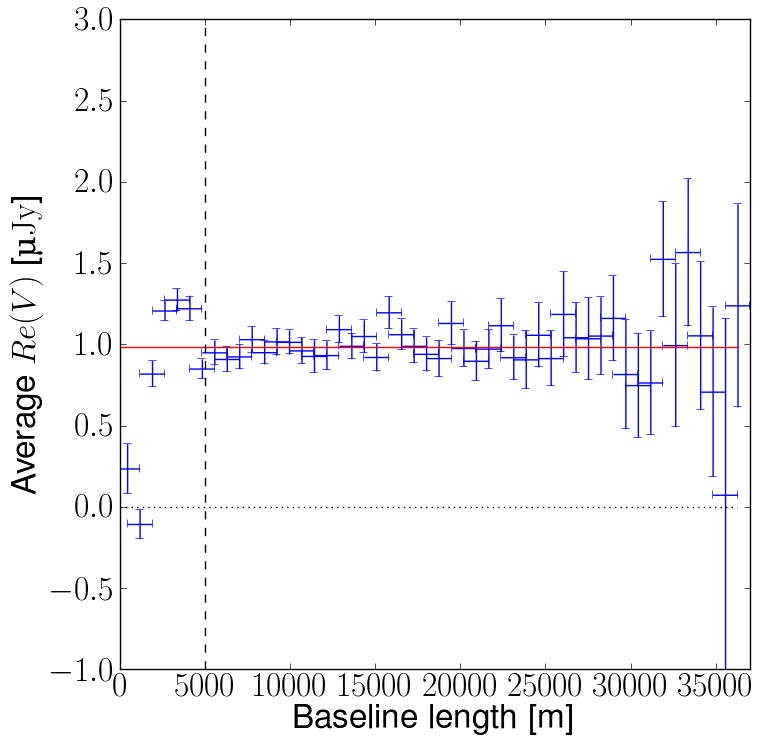} \\
\includegraphics[width=8cm,height=8cm]{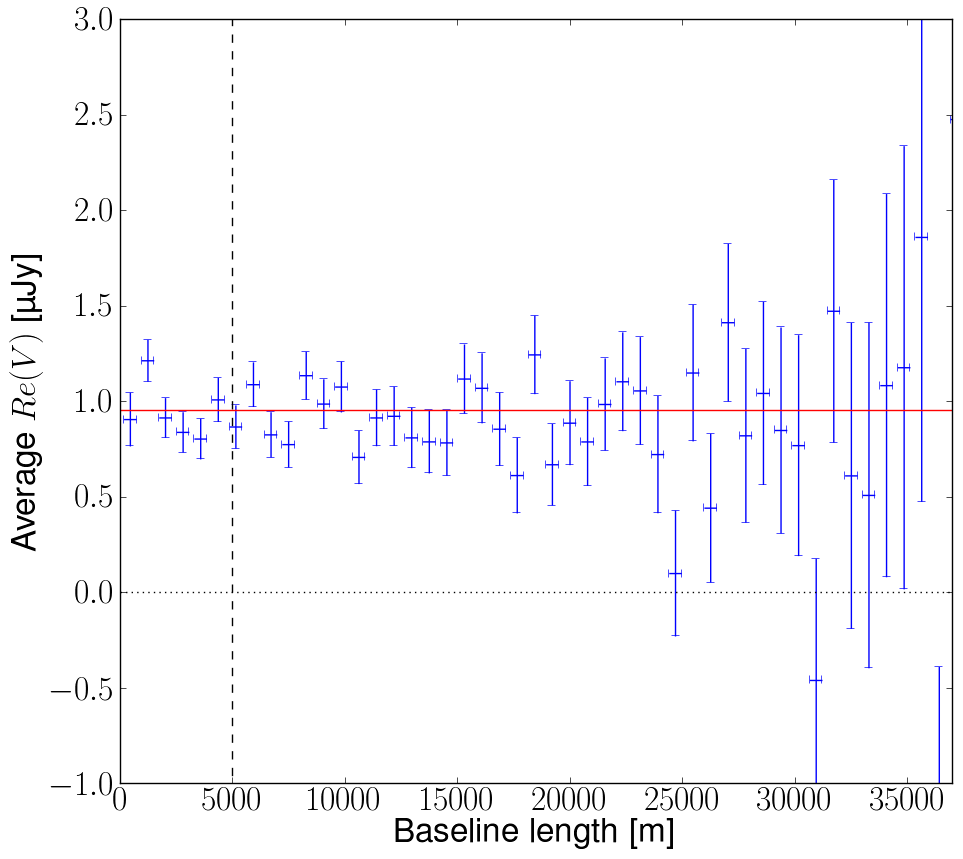}
\end{center}
\caption[]{Simulations, representative of a JVLA A-array setup at 1.4\,GHz
Top: including bright extended source, which have been removed using clean
from the data prior to stacking. Included here are the results of 100 Monte
Carlo simulations.   Bottom:  Similar simulation, however without any bright
foreground sources;  50 MC realisations.  
The artefacts seen at the short baselines, $<5000$, are most likely caused by
the imperfect removal of the bright, extended foreground sources. In real
data, the removal of the bright sources cannot be controlled beyond our best
knowledge, and therefore having the uv-stacked data available means that we
can select which baselines we use in the analysis of e.g. average flux. 
The red line represent the best fit for point source flux with the short
baselines $<5000$ are excluded. 
From \citep{lindroos15}.
\label{fig:brightforeg}}
\end{figure}

As an example, in Fig.\ \ref{fig:brightforeg}, we show the amplitude as function of
uv-distance for a uv-stacked source from a simulation with extended
bright foreground sources;  the bright foreground sources were
removed from the simulated data using the Clean algorithm. 
We find
that the short baselines of the stacked source suffer from imperfect bright
source removal.  Because the stacking was performed in the uv-plane we were
able to measure the stacked flux reliably when selecting the baselines not or
less affected by the bright source residuals, and thus determine the
stacked flux to the level of the input model.  For comparison, the fluxes
measured in with image-stacking were about 10 per cent lower than the input
model.    As a check, we have carried out a similar simulation but without
the bright foreground sources, and the resulting stacked source has no
visible problems at the short baselines.  

\subsection{Extended target sources}

\begin{figure}
\begin{center}
\includegraphics[width=8cm,height=8cm]{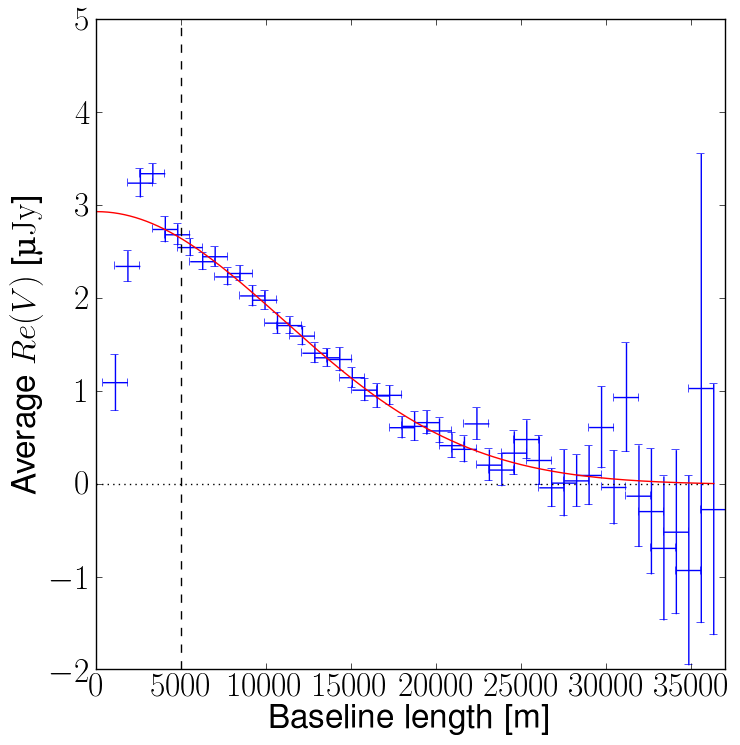}
\end{center}
\caption[]{Simulations, representative of a JVLA A-array setup at a frequency
of 1.4\,GHz. Similar to Fig.~\ref{fig:brightforeg}, top panel, however, here 
the stacked target sources have extended emission of $\sim 1.5''$.  
The red line represent the best fit for an extended source with the short
baselines $<5000$ excluded. 
From \citep{lindroos15}.  
\label{fig:extsrc}}
\end{figure}

Expanding the simulations, 
we have simulated target sources that are slightly extended with a size of
1.5'' for the JVLA A-array configuration.  
The aim of this was to determine the performance of {\it uv}-stacking vs
image-stacking in terms of estimating an average size.  
In Fig.\ \ref{fig:extsrc}, we show
the amplitude as function of uv-distance for the stacked source.  
Again, as discussed above, the shortest baselines suffer from imperfect
removal of the bright sources in the field.   
The uncertainty of size-estimate done using
the uv-stacked data is half that of the same estimate done with image-stacked data.  

The uv-stacking has the advantage of allowing for the possibility to
ignore the shortest baselines.  
In comparison the image-stacked source could be subject to the high dynamic range
issues, e.g. this could result in the presence of an extended component.  It is
possible to deal with this to some extent by using a low-spatial-frequency
filter, however, a more direct way would be to do this directly on the
uv-data. 

Additionally, in comparison with the uv-stacked source, the image-stacked
source is convolved with the beam, therefore it is best fitted with a model of
the convolved dirty beam.   
Many large surveys currently are done as mosaics with several pointings.
This means that the beam will not be the same for each pointing, and this is
a complication for image-stacking and e.g. modelling a dirty beam.

\begin{figure}
\begin{center}
\includegraphics[width=10cm,height=10cm]{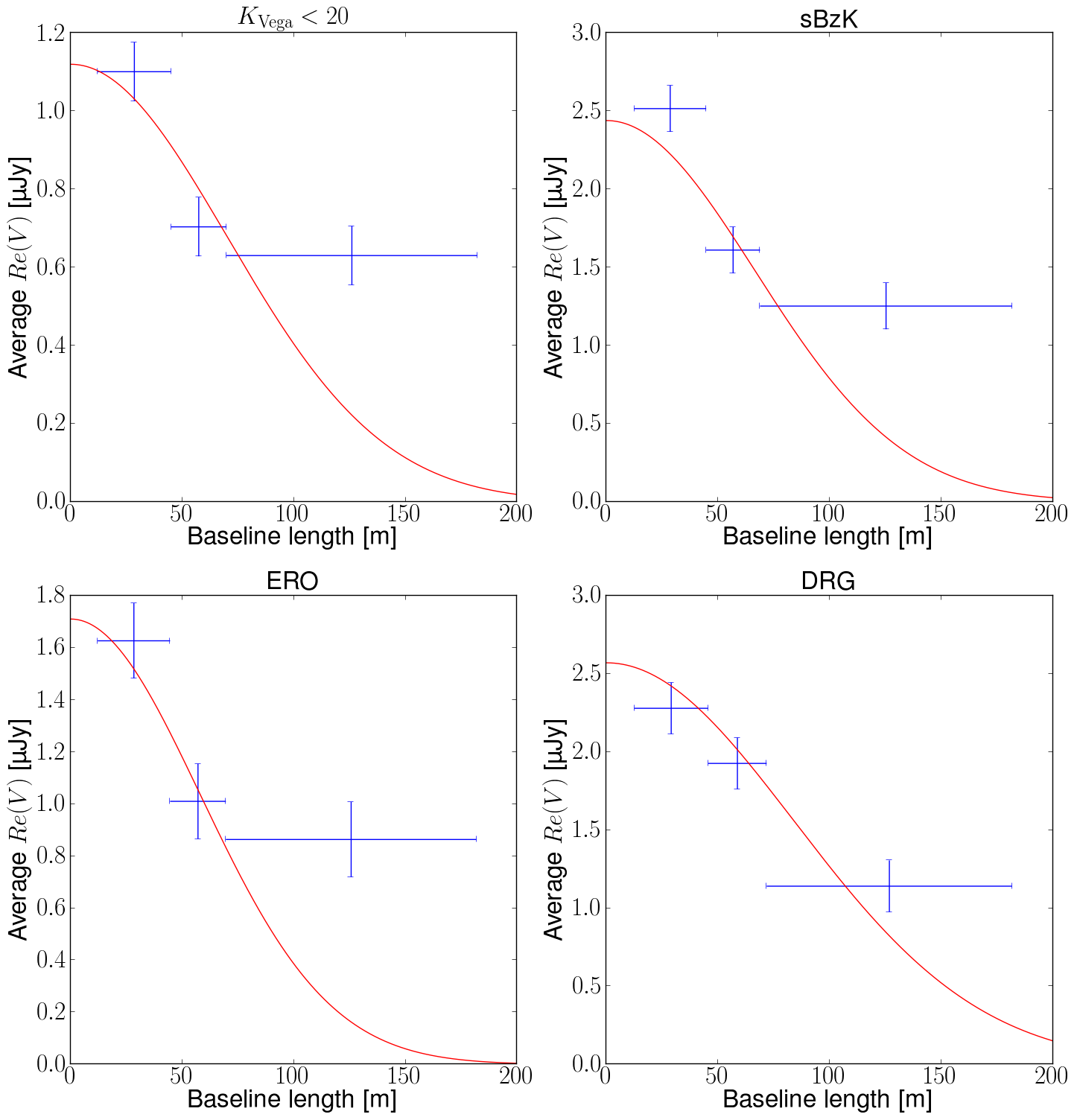}
\end{center}
\caption[]{Stacking of ALMA data for high-$z$ galaxies using four different
selection criteria (following that of \citealt{decarli14}), $K<20$, sBzK,
EROs, and DRGs.  For each uv-stacked data set, we plot the amplitude vs
baseline length (binned).  
From \citet{lindrooslic}, where similar results for VLA data are found.  
\label{fig:srcsizes}}
\end{figure}

We have applied the stacking algorithm to real data.  As presented in
\citet{lindrooslic}, 
using both VLA and ALMA data from the Extended
Chandra-Deep-Field South (VLA data: \citealt{miller13}, ALMA data:
\citealt{hodge13}), initial results yield source sizes of about $1''$ for
optically/near-infrared selected high-$z$ galaxies.  More
importantly is, however, the fact that we can use the stacked uv-data for
further interpretation.  In Fig.~\ref{fig:srcsizes}, we show the averaged
amplitude vs baseline length, and find that there is a rise towards the
shortest baselines possibly caused by an extended component, and a plateau
towards the longer baselines possibly indicating a point-like component.  
Among the possible interpretations of this, is that the stacked sources
represent a mixture of extended and compact galaxies, or that the
star-forming regions are compact but distributed over an extended region.  We
show this here, not to draw immediate conclusions about the distribution of
the radio or mm emission from high-$z$, star-forming galaxies, but to
illustrate the vast potential for further analysis of for example structural
parameters when having the stacked uv-data available.   

\section{Discussion:  Consequences for stacking of SKA data}

The tests of our new stacking algorithm, shows that the uv-stacking provides
a more robust result relative to the image-stacking.  The algorithm is tested
both in the GHz range, where currently the JVLA is the most sensitive array, and
in the 100's GHz range, where ALMA is the most sensitive array.  
This means that our findings are applicable to the SKA-MID design and
frequency range.  
The
algorithm is designed to be applicable to radio interferometry in general, so
also for lower frequencies such as SKA-LOW.
The access to the stacked uv-data provides possibilities for a more robust
analysis, in particular it enables reliable filtering at different spatial
frequencies as well as means for detailed analysis of the stacked sizes.  

Stacking is mentioned by many different future projects for the SKA, and
therefore it is important to make available the tools that provide the most
optimized stacking tools.  We argue that uv-stacking is such a powerful tool,
that it should be standard for future facilities.  
However, for future facilities such as the SKA, this also means facing a 
significant challenge, as it is
expected that most raw and even calibrated SKA uv-data will not be archived,
at least not for long-term storage.  Consequently, one could argue that
uv-data should also be stored long-term, unless the
stacking is restricted to the image plane. In that case one needs to be aware 
that the results are less robust and more prone to additional uncertainties.

We consider three different options: 
\begin{enumerate}
\item Design a {\it stacking queue}, which will be executed in almost real-time
with the observations.  The uv-stacking would be carried out after
calibration, during a 'buffering' time before the uv-data is
removed.  In the case where the processed and calibrated uv-data will not be
stored for long term archive access, but only kept for a short period needed
for processing of (large) surveys, one could imaging that the SKA should
offer an 'observing mode' where stacking lists are submitted, and then
processed in parallel with the rest of the survey.  This compromise would
enable the stacked uv-data to be available for readily defined positions, as such a data set will be significantly reduced in size.  The Lindroos et al (2015)
algorithm is designed so that it could easily be tailored to such
a purpose.  
\item Ensure that calibrated uv-data is archived, if not all, then at least in
some averaged format and at least for large surveys.   This would enable the
most flexible processing of the data in terms of stacking, also
for astronomers not directly involved in the surveys.  
\item Only stacking in the image-plane:  Accepting potentially less realible
and less robust results in favour of not have to archive uv-data.  However, this
also means limitation for a number of different aspects, e.g. measuring sizes
of stacked sources. Probably most importantly, it would mean no access to baseline
info and thus limited possibility to filter out short baselines as
illustrated in Figs.\ \ref{fig:brightforeg} and \ref{fig:extsrc}.

\end{enumerate}


\section{Summary}

In this chapter we have discussed stacking of uv-data vs image-data from the
perspective of the SKA.  We describe our novel algorithm for stacking of
visibilities, which we have compared to image-stacking.  The uv-stacking algorithm
has been developed for application to any type of radio interferometric data. 
The comparison was done
primarily using simulated data based on JVLA and ALMA.    

We have found in a detailed comparison that
uv-stacking is a more robust method than image-stacking.  The results
produced from uv-stacking are either similar or more reliable than the
image-stacked results.  Having access to the stacked uv-data provides means
it will be possible to remove or at least treat artefacts such as imperfect removal of bright
sources, which can signficantly affect the stack results.  Furthermore,
having the stacked uv-data in hand enables a more detailed analysis of the
properties of the stacked source, in particular size measurements.  

It is therefore our conclusion that in the design of the SKA it should be
carefully considered to allow for uv-stacking, either through the
means of making (averaged) calibrated uv-data available in archives (at
least for selected surveys), or as an alternative 
provide a specially designed {\it stacking queue} for large/deep surveys.  
Having access to the uv-data after stacking will be invaluable in ensuring
that the desired signal can be optimally extracted from the data.

\bibliographystyle{apj}

\end{document}